\documentstyle[referee]{l-aa} 

\begin{document}
\thesaurus{03      		 
              (02.18.7; 	 
	      11.01.2;		 
   	      11.03.3;		 
	      11.09.4)}		 

\title{Ca depletion and the presence of dust in large scale
nebulosities in radiogalaxies (I)}


\author{
M.Villar-Mart\'\i n
\inst{1}
\inst{2}
\and
L.Binette
\inst{3}
\inst{4}
}

\institute{$^{1}$\,ST-ECF, Karl-Schwarschild-Str 2, D-85748 Garching, Germany
\\
$^{2}$\,Max-Planck-Insitute f\"ur extraterrestrische Physik,
Giessenbachstrasse,
Postfach 1603, D-85740 Garching, Germany
$^{3}$\,Observatoire de Lyon, UMR CNRS 142, 9 av. Charles Andr\'e, F-69561
Saint-Genis-Laval Cedex,
France  \\
$^{4}$\,Space Telescope Science Institute, 3700 San Martin Drive, Baltimore,
Maryland
21218, USA}

\date{Received  1995 June 6; accepted 1995 Sept 12 }

\maketitle

\begin{abstract}
	We show that the study of the Calcium depletion is a valid and
highly sensitive method for investigating the chemical and physical
history of the very extended ionized nebulae seen around radio
galaxies (EELR), massive ellipticals and `cooling flow' galaxies.  By
observing the near IR spectrum of nebular regions characterized by low
excitation emission lines (LINER-like), we can use the intensity of
the [CaII]$\lambda\lambda 7291,7324$\AA\ doublet --relative to other
lines, like H$\alpha$-- to infer the amount of Calcium depletion onto
dust grains. The presence of dust in these objects --which does not
necessarily result in a measurable level of extinction-- would favour
a `galactic debris' rather than a `cooling flow' origin for the
emitting gas. Before aplying such test to our data, we study four
possible alternative mechanisms to dust depletion and which could have
explained
the absence of the [CaII] lines: a) ionization of Ca$^+$ from its
metastable level, b) thermal ionization of Ca$^+$, c) a high
ionization parameter and/or a harder ionizing contiuum than usually
asummed and d) matter bounded models associated to a hard ionizing continuum.
We show that none of these alternative mechanisms explain the
absence of the [CaII] lines, except possibly for the highly ionized
EELR where  a high ionization parameter is required combined with a soft power
law. We thus conclude that for the other low
excitation emission regions (cooling flows, liners, low excitation EELR), the
abscence of the CaII lines {\it must} be due to the depletion of Calcium onto
dust grains.

\end{abstract}

\section{Introduction}

The study of the interstellar medium (ISM) of external galaxies
provides important information about the global kinematic (inflow,
outflow) of such gas, its chemical composition and the implied star
formation history, its mass distribution, etc.  This gas forms a vital
part of the record of the formation of the parent galaxy, and the
evolutionary processes involved.

	How can this material be studied in details? One way is to
have it illuminated or excited by a powerfull AGN, gi\-ving rise to the
phenomenon of extended emission line region (hereafter EELR). The
drawback of course is that it only allows us to look at a restricted
class of galaxy. The large scale EELR phenomenon is observed in a
majority of the most powerful radiogalaxies with EELR extending to
radial distances of up to 100 kpc from the nuclei (Tadhunter 1986;
Baum et al. 1988), much larger radii than the stellar population
distribution of the parent galaxy. The morphologies and kinematics of
such regions cover the full range from regular disc/ring systems to
chaotic systems for which no pattern can be discerned.  Their spectrum
show strong emission lines, covering a wide range in ionization.  It
is generally accepted that the EELR are ionized by some mechanism
connected with the nuclear activity, but there is no full concensus on
the excitation mechanism.  Some objects show evidences for an
interaction between the radio jets, which transport ener\-gy to the
outer radio lobes, and the gas in the outer region. Maybe this
interaction is responsible of the excitation of the gas through some
kind of shocks (Sutherland, Bicknell \& Dopita 1993).  For other
objects, that do not show evident spatial coincidence between the
radio structures and the EELR, the excitation might be due to direct
photoionization by the nuclear ionizing radiation field (Robinson
et{}~al. 1987: hereafter RBFT87). A reduced scale version of the EELR is
the one observed in many Seyfert galaxies (Haniff, Ward \& Wilson
1988) where the ionized gas may extend up to a few kpc although it is
brightest within the central 100--300pc. The EELR morphology tends to
be conical (Wilson \& Tsvetanov 1994). Its detailed observation is
complicated by presence of the very luminous stellar background of the
bulge. In normal bright ellipticals, quite weak extended nebulosities
of low excitation is a common phenomenon (Buson et{}~al 1994, Goodfroij
1994). The very large scale gas around radio-galaxies and on which we
focus here presents the advantage that the lines are observed against
the sky background rather than against the bright parent galaxy bulge.

	The origin of the gas making up these {\it large scale}
extended nebulosities remains unknown. Furthermore, the distinction
between these  and the filamentary ne\-bulae seen in some clusters and around
some massive ellipticals and often identified with cooling flows, is
unclear.  We do know, however, that this gas in every case is
chemically enriched as compared to primordial gas. Emission line
analyses (RBFT87) show common element abundances to be
within a factor of a few ($\le$) of Solar and also to be rather
uniform over all the objects observed.

	The two most likely explanations for the origin of the material  are:

	a) debris from recent tidal interactions and mergers

	b) gas cooling from the hot ($\sim$ virial) phase which from X-ray
observations (Forman, Jones \& Tucker 1985) has been shown to exist around
massive
ellipticals and inside galaxy clusters

The main arguments behind these explanations are the following:

	a) Heckman et{}~al. (1986) showed that a large fraction of powerful
radiogalaxies
have morphological features --shells, tails, loops, etc-- similar to those
produced in
numerical si\-mulations of galaxy interactions (e.g., Toomre and Toomre 1972,
Quinn
1984). This could indicate that the activity has been triggered either because
fresh
gas has been accreted from outside or because prexisting gas in the galaxy has
been
caused to collapse to the core as a result of the interaction. This interaction
scenario is also supported by observations of a few nearby radio   galaxies
which,
apart from morphological peculiarities, show large misalignments between the
stellar
and the gaseous rotation axes, indicative of an external origin for the gas.

 	b) Hot X-ray coronae ($T \ge 10^4$K) are a common feature of bright
early-type
galaxies.  Within some critical radius, radiative cooling becomes important,
leading
to the deve\-lopment of the so-called `cooling flow' hypothesis (Nulsen,
Stewart \&
Fabian 1984,Thomas 1986, Thomas et{}~al. 1986) Eventually, condensations or
filaments
could be formed, dense and cool enough to radiate detectable optical emission
lines.
Most powerful radio galaxies are too distant for the characteristic X-ray
emission to
be currently detectable, but it is quite plausible that the EELR gas has
condensed
out of a surrounding cooling flow.  Any discovery of their EELR as consisting
of
extensive optical filaments would argue in favour of the cooling flow
hypothesis.

	Distinguishing between these alternative hypotheses has been attempted using
gas kinematic measurements (Tadhunter, Fosbury \& Quinn 1989) which show that
the
radio galaxy EELR generally have a high specific angular momentum which is
difficult
to reconcile with the cooling flow picture. An alternative approach is to look
for
the presence of dust associated with the EELR gas. If the gas has cooled
directly
from a hot phase, there will have been no opportunity for dust to form,
according to
the standard cooling flow theory. Any dust introduced from galaxies into the
hot
intracluster medium will be rapidly destroyed (Draine \& Salpeter 1979). We
would
need a mechanism to produce this dust once the gas has cooled down (Fabian,
Johnstone
and Daines 1994).  If, on the other hand, the material has fallen in during a
merger,
the dust/gas ratio is expected to have a value appropriate to the gas chemical
composition found in normal galaxies.

Determining the presence or absence of dust is important because:

-of its relevance in deriving chemical composition which takes into account the
effects of depletion of metals unto dust grains

-of the implication for the star formation history and the ISM evolution: when
does
dust form? (and so stars?)

-of the effects of dust on the apparent morphology of continuum and line
features:
pure absorption (reddening) and/or scattering (blueing/polarization)

We discuss below how the absence of the forbidden [CaII] lines can be
used to infer the presence of dust mixed in with the emission gas. We
assess in detail all the most plausible {\it alternative} explanations
to that of internal dust for explaining the absence of [CaII]
lines. As no acceptable alternative solution is found, we conclude in
favour of the validity of the method initially proposed by Ferland
(1992). We however adapt and optimize the [CaII] dust detection
technique to the context of the EELR studies in which we are
involved. The observational results and their interpretation will be
presented in a subsequent paper.

\section{Outline of the method}

To investigate how the forbidden Calcium doublet of [CaII] in the
infrared is affected by the physical condition encountered in
photoionized plasma, we outline first the computer code which we used
and then proceed to illustrate how the [CaII] lines might be used to
infer dust in the ISM of galaxies and thus our interest in securing
these conclusions by closing up the possibility of alternative
interpretations.

\subsection{The photoionization code MAPPINGS.}

	To compute the emission lines used in our study, we have used
the multipurpose photoionization--shock code MAPPINGS (c.f., Binette,
Dopita \& Tuothy 1985). The Ca$^+$ ion is considered as a five level
atom. Its structure is shown in Fig.{}~1.

	One interesting aspect of the updated (Binette et{}~al 1993a,b)
code MAPPINGS is that the effect of dust scattering on the line
transfer is explicitly solved using the numerical solution of Bruzual,
Magris \& Calvet (1988) as described in Appendix{}~A of Binette et{}~al
1993b.  Other effects of dust on the ionization structure as well as
on the thermal balance of the plasma are considered followoing the
prescriptions of Baldwin et{}~al. (1991 see Appendix{}~C). The dust grain
charge is calculated self-consistently and the formula describing the
photoelectron energy distribution and the yield are from Draine (1978)
but with a cap of 0.2 for the yield at high photon energies.

	We now proceed to define the parameters employed in the
calculations. Most of these are derived from our observational
knowledge of EELR although we recognize that it is incomplete and
can be biased by particular diagnostic tools which are employed.

\subsubsection{Metallicity}

For definiteness, we adopt a set of solar abundances (Anders \&
Grevesse 1989) for the trace elements, in line with the finding of
RBFT87 who indicated values for the EELR not radically
different from solar. The solar abundance of ({\it not} depleted) atomic
Calcium is $2.2 \times 10^{-6}$ by number relative to
hydrogen. Depletion into interstellar dust grains is known to reduce
this abundance in the local ISM by a factor of $\sim 5000$ (Whittet
1992) so that if depletion was taking place in the partially ionized
EELR plasma, even a small dust-to-gas ratio might be sufficient to
eliminate any detectable trace of atomic Calcium.

\subsubsection{Geometry}

	It's been accepted for a long time that the NLR and the EELR
are formed by individual clouds that are ionized by the central
source.  In most models presented here, each emitting cloud is
considered as a radiation bounded slab (optically thick to the Lyman
continuum) which comprises: (1) a fully ionized region, closer to the
illuminated face and responsible for the high ionization lines; (2) a
partially ionized zone (PIZ), where the low ionization lines like CaII
are emitted.  The boundary of the photoexcited regions (1 + 2) is
defined as the depth where the following two conditions are
simultaneously satisfied: a) The unabsorbed ionizing flux is $< 1$ \%
of the impinging flux and b) the ionized fraction $n_{H^+}/n_H\leq 1$
\%. Matter bounded clouds will be considered in Sect.{}~3.4. As we want
to concentrate on the depletion phenomena, we will simplify the
calculations by not considering any neutral region (3, see Fig.{}~2)
beyond the PIZ which would contain dust, the effect of which would be
to cause additional extinction for an hypothetical and unfavorably
placed observer looking from the back of the slab. We have verified
that the results reached here using line ratios of similar wavelengths
are not altered by the presence or not of this neutral absorbing zone.

\subsubsection{Gas density}

	We adopt a representative density for the EELR clouds of 300
cm$^{-3}$. Typical electron densities values measured vary between a
few tens (or less) to a few hundreds.  The densities derived from
forbidden line ratios might not apply to every subregion, but they are
sufficietly low to consider the low density limit as a generally valid
and very good approximation to the physical conditions affecting the
CaII lines of large scale EELR. In the low density regime, the effects
of density variations on the emission line ratios which are considered
in this paper are very small by comparison to the effects of other
parameters like the ionization parameter or the hardness of the
ionizing continuum.

The calculations consider the gas pressure to be cons\-tant within the
cloud (isobaric models) with the density behaviour modulated with
depth into the cloud by the behaviour of the temperature and by the
ionization fraction of the gas.

\subsubsection{The ionizing continuum and the ionization parameter U}

	We implicitely assume that the dominant ionization me\-chanism
of the ionized gas in radio galaxies is photoionization. It has been
shown by RBFT87 that a hard continuum extending well
down into the soft X-ray region, despite some discrepancies with the
observed spectra, can be considered to reproduce generally well the
measured line ratios. This continuum can take its source in the active
nucleus or be locally generated (e.g., fast shocks: Binette, Dopita,
Tuohy 1985). We have considered power law (PL) distributions of various
values of $\alpha$ ($F_\nu \propto \nu^{+\alpha}$) as well as hot
blackbodies (BB) of temperature T$_{bb} \approx 10^{5}K$ in order to study
the effects that hardness has on our conclusions.

	The ionization parameter, a measure of the excitation level of
the ionized gas, is defined as the quotient between the density of
impinging ionizing photons and the density of the gas cloud:

U$ = \frac{Ionizing \, energy \, flux}{c \, n_H} =
\frac{\int{L_{\nu}d\nu/h\nu}}{4\pi r^2cn_H}$

\vspace{0.2cm}

\noindent where $L_{\nu}$ is the monochromatic luminosity of the source,
r the distance of the cloud to the continuum source, $n_H$ the density of
the gas and $c$ the speed of the light.

	We find that the parameters having the strongest effect on the
line spectrum are the ionization parameter U and the mean ionizing
photon energy ({\it i.e.,} hardness of the continuum).

\subsection{How to detect dust}

	How can we detect dust within the gas associated with EELR? We here
summarize different techniques and compare their sensitivities.

	a) Reddenning: given the nature of the gas distribution and
the fact that it is ionized externally (unlike HII
regions which are internally excited), the extinction may not be
necessarily sufficient high to be easily detected using optical
observations. Indeed, the EELR ratios studied by RBFT87 show little or
no reddening.  Furthermore a small enhancement of the Balmer decrement
over recombination case{}~B might be interpreted as resulting from
collisional excitation rather than from reddening.

	b) Scattering: polarization measurements of high redshift
radio galaxies (di Serego Alighieri {\it et al.} 1989, 1992, Januzi \&
Elston 1991, Tadhunter {\it et{}~al.} 1992) show conclusive evidence
for scattered nuclear light over large volumes and, altough there are
reasons to believe that the scattering medium is dust, it is difficult
to rule out entirely Thomson scattering by hot electrons. Detailed
studies of the low redshift galaxy PKS2152-69 do, however, show
polarized continuum radiation from highly excited extranuclear gas
cloud with an energy distribution which is so blue that it must arise
from dust scattering (Fosbury {\it et al.} 1990).

	c) Infra-red thermal dust emission: the IRAS satellite has
shown that many galaxies radiate significant fractions of their energy
in the far infrared sprectral region. Significant masses of dust at
temperatures of around 40K are responsible of this radiation at
wavelengths of 60$\mu$m and beyond. In many cases the FIR spectral
energy distributions is still rising at 100$\mu$m, out of the spectral
range detectable by IRAS. The cool dust can only be detected at
milimetre and submilimetre wavelengths. A strong limitation of this
technique is the very poor spatial resolution of the IRAS satellite.
Groundbased studies of the far IR emission of galaxies in the sub-mm
range also exist({\it e.g.} Clements, Andreani \& Chase 1993), but
still with poor spatial resolution.  Although it is possible to infer
masses and temperatures of the warm and cool dust components
(dependent on models), the spatial distribution of the dust is not
known.

	d) Indirect effects on the line spectrum. There are se\-veral
ways this can happen: effects of dust on the gas temperature
(photoionization of dust grains may raise the temperature of the
plasma), effects of dust on the ionisation structure (dust grains selectively
absorb ionising photons of lower energies), and influen\-ce on apparent
chemical
composition via the depletion of refractory elements onto dust grains.
The first effect does not provide a unique interpretation for the
unusually high temperatures seen in some EELR (Tadhunter, Robinson \& Morganti
1989) while the second effect cannot be discriminated
against reliably since even dust-free photoionization models are still
too uncertain to be used as absolute reference point. For these
reasons, the last effect is the only clearly promising one and is
looked into details below.

Calcium is very sensitive to the presence of dust as it is always
found to be depleted in the interstellar medium (e.g., Crinklaw,
Federman \& Joseph 1994).  Photoionization calculations appropriate
to LINERS --a hard ioni\-zing spectrum with a relatively low
ionization parameter-- invariably predict the [CaII]$\lambda\lambda
7291,7324$\AA\ doublet (4$s^2$S-3$d^2$D) to be very strong (Ferland
1993).  These two forbidden lines (hereafter
F1$\equiv\lambda7291$\AA\, F2$\equiv\lambda7324$\AA\ ) have a high
critical density $\sim 10^6$ cm$^{-3}$. The latter line, 7324\AA\, is
the weakest of the doublet and is furthermore blended with the
[OII]$\lambda7325$\AA\ multiplet. The other line, $\lambda7291$\AA\,
lies some 30\AA\ shortward of [OII] and is therefore straighforward to
isolate given reasonable spectral resolution.  The fact that any of
these doublet lines are generally not seen in \-LINERS but are so in
some novae when their envelopes have reached the appropiate ionization
level can be interpreted as evidence of Calcium depletion onto dust
grains in the former objects. Since EELR which are the subject of our
investigation are often seen in the ionization parameter region of the
line ratio diagnostic diagrams occupied by LINERS ($10^{-4}\leq U\leq
10^{-3}$), we can similarly use the [CaII] doublet measurements to
infer whether or not there is depletion taking place in EELRs and thus
conclude whether dust is also mixed with the gas as is thought to be
the case in LINERs.

	 The above arguments have already been used for se\-veral
objects with `cooling flow' filaments by Donahue \& Voit (1993) to
infer the presence of dust mixed with the ionized gas. Ferland (1993)
has shown the great sensitivity of this method to the presence of dust
under NLR conditions.  We now show it to be also the case under EELR
conditions. We present two diagnostic diagrams in Fig.{}~3, the ratio
[CaII]/H$\alpha$ (7291/6563) and the ratio [CaII]/[ArIII] (7291/7135)
as a function of the ionization parameter $U$.  The variable parameter
distinguishing the three different sequences in U is $\mu$, the
dust-to-gas ratio of the plasma expressed in units of the solar
neighborhood dust-to-gas ratio. The dramatic difference in line ratios
between the grain depleted Ca/H ($\mu > 0$) and the undepleted case
($\mu = 0$), shows the sensitivity of this method, particularly for
low values of $U$.


\section{Possible alternative explanations to depletion.}

	 Before we carry the conclusions of the current analysis to
the interpretation of our observations (Villar-Mart\'\i n \& Binette
1995), we report first on our effort in investigating other possible
alternative mechanisms to dust depletion. If the warm Ca$^+$ region
predicted by standard models does exist, then Calcium depletion
becomes the only reasonable explanation for the non detection of the
doublet lines. What we consider in this section is the possible NON
EXISTENCE of the emitting [CaII] region by investigating different
mechanisms which could eliminate it. During
our investigation, we require however that successfull models do not result in
important
discrepancies with other observed line ratios. The mechanisms we have
considered to eliminate the [CaII] region are

	- Ionization of Ca$^{+*}$ by Ly$\alpha$ and soft continuum photons from the
metastable level of Ca$^+$

	- Thermal (collisional) ionization of Ca$^+$

	- Photoionization with a much harder continuum or a much
higher U than usually asummed

\subsection{Ionization by Ly$\alpha$ and soft continuum photons.}

	Wyse (1941) proposed that the ionization of CaII from the
me\-tastable level by Ly$\alpha$ photons, could explain the fact that
the IR lines of CaII at 8498, 8542 and 8662\AA\ appear in emission
near the maximum phase of Me variables, whereas the H and K lines only
occur in absorption.  Trapped Ly$\alpha$ photons could also play a
part in ionizing metastable Ca$^{+*}$ as suggested by Wallerstein
et{}~al. (1986).

	We investigate here if this process is important under the
conditions found in EELR clouds. In order to do this, we add two terms
to the ionization equilibrium equation of CaII. One which considers
photoionization of excited Ca$^{+*}$ by the impinging UV continuum.
The other is photoionization of excited Ca$^{+*}$ by the nebular
Ly$\alpha$ photons.  The statistical equilibrium equations give the
relative population of the mestastable level, which turns out to be,
under EELR conditions, $\frac{n_{3d}}{n_{4s}}\sim 10^{-7}$, being
$n_{4s}$ the density of Ca$^+$ ions in the ground level. With such a
negligible population, the density of ionizing photons must be very
high to increase the ionization rate to a non negligible level as
compared to the ground state ionization rate.  A simple estimate
presented in Appendix demonstrates this level to be out of reach.

	In summary, the very diluted radiation fields and the low
densities appropriate to the EELR implies an extremely small
population for the excited levels which prevents the ionization of
Ca$^{+*}$ by Ly$\alpha$ and soft continuum photons from being of any
significance.

\subsection{Thermal ionization of Ca$^+$.}

	 We investigate here the possibility of collisional ionization
by thermal electrons of Ca$^+$ to Ca$^{++}$, a process which is
important when the electronic temperature becomes higher than 20000K
(Jordan 1969). In order to establish a comparison in U, we have
considered two extreme cases in our calculations, log$U=-4{}~ \& -2$. To
illustrate how a much harder continuum will result in much higher gas
temperatures, we also use two different PL of index $\alpha -1.4$ and
$-0.4$.  Note that such a hard continuum as $\alpha =-0.4$ is probably
quite unrealistic. However, our intention here is simply to test
whether very high temperatures can be achieved with photoionization
models and specifically near the Ca$^+$ region. The results are shown
in Fig.{}~4 as a function of depth in the photoionized slab.  Of the
eight plots, the four upper ones correspond to $\alpha = -1.4$ while
the four at the bottom to $\alpha = -0.4$. The four plots on the
right, have logU$=-2$, and the four on the left, logU$=-4$. Two plots
therefore are shown for each pair of [U, $\alpha$] values: one is the
the temperature T4 in units of 10000K and the one immediately
underneath is the intensity of F1 ($erg.s^{-1}.cm^{-2}$), both as a
function of the depth in units of $10^{20} cm$ into the slab. These
plots allow us to see the correspondig electron temperature to the
position where the bulk of the [CaII]F1 emission takes place.

	For the traditional PL of index $\alpha=-1.4$, the electron
temperatures are not anywhere near high enough for the process of
thermal ionization to be relevant. Harder continua and increasing U
values do produce higher temperatures, but, unless U are
unrealistically high (even logU$=-2$ is not enough), the gas is never
hot enough. Thus, realistic photoionization models are {\it not}
able to heat the gas sufficiently to thermally ionize Ca$^+$. We might
conjecture that there could be an additional heating source like
shocks which could raise the temperature of the gas. This would be an
interesting point for further investigation.

\subsection{Effects on F1 of varying U and the continuum hardness. }

	We now investigate how the F1 line might become undetectable
by simply varying arbitrarily the ionization para\-meter or the
continuum hardness. Fig.{}~5 shows six diagnostic diagrams with the absciss
always representing the line ratio \-
[OI]$\lambda6300$/[OIII]$\lambda5007$. The [OI]/[OIII] ratio
monotonically increases with
decreasing gas excitation (i.e., with decreasing U) and is therefore
a good measure of the
excitation level of the gas. Each dia\-gram
shows in ordinate a different line ratio which can be related to a
given gas property. [OI]$\lambda6300$/H$\alpha$, for instance, might
measure the hardness of the continuum. In the last diagram, the
ordinate corresponds to the quotient F1/[OI]$\lambda6300$.  The three
sequences of models shown in each diagram differ by the slope of the
power law which takes on the values of $\alpha =-1.4, -1$ and $-0.4$. The
values of Log{}~U covered by each curve is in the range $-4$ to $-1$.
Our aim is to look for models which can decrease the F1 intensity
below the detection limit. Let's look at how we might define a
practical detection limit. The open squares in the diagrams of Fig.{}~5
represent line ratios measured by RBFT87 in several EELR. The faintest
line they measure is typically HeII$\lambda$4686. The mean ratio of
HeII$\lambda$4686/[OI]$\lambda$6300 observed is $10^{-0.4}=0.4$ for the
large scale nebulosities. We establish our 'artificial' detection
limit in the following way: since HeII is one of the weakest line
successfully measured by RBFT87, we will assume that any line fainter
than 0.4 below the [OI]$\lambda$6300 flux is not detectable. In
the last diagram, the region where F1 falls below this detection limit
is shown by a dash line grid. Any model found  in this area is deemed
successful
in explaining the non-detection of F1 without requiring depletion.

	We see that for the standard PL ($\alpha=-1.4$), only mo\-dels
with high U (log U$>-2$) decrease log(F1/[OI]) below --0.4.  These
models, as we can see in the diagrams, would therefore be valid only
for the high excitation EELR, but not for LINERs, cooling flow
filaments, or EELR of low and intermediate excitation. On the other
hand, increasing the hardness of the continuum (flatter power laws),
helps F1 to get fainter with respect to [OI]$\lambda$6300, but the
discrepancies with observed line ratios in other diagrams become
totally unacceptable (see top two diagrams of Fig.{}~5).

	It is interesting to compare a BB sequence ($1.2\times
10^5$K) with the canonical PL sequence $\alpha = -1.4$. We see that
both ionizing continua reproduce rather well the observed line ratios
as was earlier shown by RBFT87. From these line ratios alone, there
are no reasons to favour power laws over hot blackbodies. A similar
conclusion was reached by Binette, Robinson and Courvoisier (1988) for
the mean NLR spectrum of Seyferts.

	A BB produces a much stronger F1 compared to [OI]$\lambda$6300
than any of the power laws considered.  One reason for this is that
the fraction O$^o$/O$^+$ in the PIZ is completely controlled by the
charge exchange reactions of O$^o$ and O$^+$ with H$^+$ and H$^0$,
respectively, and not by direct photoionization of O$^o$. This is not
so for Ca$^+$/Ca$^o$ which is free to respond to the different amount
of hard photons (the only one to make it to the PIZ) available in a PL
or a BB.  We conclude from the last diagnostic diagram that the BB
models could not explain the abscence of F1 from the observed spectra
without invoking depletion.

\subsection{Effects on F1 of truncating clouds }

        We showed in the previous section that a harder continuum can
potentially bring F1 under the detection limit but result in important
discrepancies with the low excitation lines. The reason is that the
partially ionized zone (PIZ) where most of the low excitation lines
are generated gets larger and larger with increasing hardness of the
continuum. If the clouds were truncated, the smaller PIZ would
generate weaker low ionization lines, thus improving the overall fit.

        We have investigated models with $\alpha=-0.4$
which were truncated at a depth which satisfies a given
criterion based on a specific line ratio. This has been done in two
ways:

\vspace{0.2cm}

\noindent 1) The criterion in this case is to truncate the calculations
when OIII/H$\beta$ has reached a value of 10 which is the typical ratio for the
high excitation EELR.

        The sequence of models shown in Fig.{}~6 are separated by a
factor of 0.14dex in U.  Altough there are still discrepancies, there
is a notable improvement compared to the radiation bounded models of
Fig.{}~5. The predicted line ratios are now located closer to the
observed data (same scale as in Fig.{}~5).  We conclude that models with
a hard ionizing continuum must be matter bounded in order to fit
acceptably most observed ratios.

        What happens now with F1/[OI]? In the last diagram of Fig{}~6, we see
that all these models produce F1 above the detection limits and cannot
therefore explain the non detection of Calcium.

\vspace{0.2cm}

\noindent 2) The second method consists in imposing that the models
produce F1 under the chosen detection limit (F1/[OI]=0.4) and check if
such models agree with the position of the observed line ratios in the
rest of the diagrams.

         The models in the sequence which satisfy this criterion are
found in the range $3.7\times 10^{-3} \leq U \leq 2.7\times 10^{-2}$.
They are represented in Fig.{}~6 as open triangles connected by a solid
line.  As we see in the top left diagram, the ratio [OIII]/H$\beta$ is
not any more defining a simple trend with excitation (which is
represented by the ratio [OI]/[OIII]).  Furthermore [OI]/H$\alpha$
remains a discrepant ratio as in Fig.{}~5. So although it is in
principle possible to satisfy the F1/[OI]=0.4 criterion with a hard
power law, the truncation must be done at a specific yet {\it ad{}~hoc}
depth and furthermore the previous trend of excitation with U has
disappeared.

\vspace{0.2cm}

Without rejecting the possibility of a more complicate mixture of
matter and radiation bounded models, we believe that simply truncating
clouds does not convincingly solve the problem of the weakness of the
CaII doublet and dust depletion remains the most likely
interpretation.

	It is interesting to note that truncated clouds adjust better the
HeII/H$\beta$ ratio (bottom left diagram) as proposed before by Morganti
{\it et al.} (1991) and Viegas and Prieto (1992).

\section{Conclusions}

	This work is based on the method proposed by Ferland (1993) to
investigate the presence of dust mixed with the gas of the Narrow Line
Region of active galaxies.  Because photoionization models predict
remarkably strong forbidden lines [CaII]$\lambda\lambda7291,7324$\AA\
assuming reasonable abundances of atomic Ca, the basic idea is to
infer a systematic depletion of Calcium onto dust grains whenever the
infrared [CaII] lines are observed very weak or undetected. This test
of the dust content was applied to cooling flow filaments by Donahue
\& Voit (1993) who concluded on the presence of dust.

	We have shown here that this sensitive method is also
applicable to the conditions found in the EELR of radiogalaxies. In
order to make more secure any inference about the presence of dust
based on [CaII] lines, we have investigated alternative explanations
for their absence: ionization of Ca$^{+*}$ to Ca$^{++}$ by Ly$\alpha$
photons and soft continuum photons from the metastable level of
Ca$^+$, thermal ionization of Ca$^+$, ionization of Ca$^+$ due to
either a very high U value (ionization bounded case) or to a hard
continuum (with truncated clouds). Except for the highly excited EELR
which might not possess any Ca$^+$ region due to their high ionization
level, the results are negative: none of the alternative mechanisms or
models studied can explain the absence of the [CaII] lines without
dust depletion.

	Our conclusion is that the dust content test appears generally valid for the
EELR of radiogalaxies (unless the excitation level of the gas is extremely
high). This will allow us to make important conclusions about the origin
of such gas, discriminating between galactic debris and the standard
cooling flow theory.

	A BB ionizing continuum characterized by a temperature of
$1.2\times 10^5$K can reproduce the observed line ratios of the EELR
at least as well as a PL of index $\alpha=-1.4$. On the other hand,
the F1/[OI] from a BB is higher than that of a PL so the case in
favour of depletion is even stronger.

	In a follow up paper, we will present long slit spectra of
EELR, cooling flow filaments and Seyfert{}~2 NLR, all taken in the
region of the [CaII] doublet.  The goal will be to apply the test of
the Calcium depletion described above in order to conclude whether or
not the gas in these nebulosities is mixed with dust. This will be our
starting point for deciphering the origin of the emitting gas.

\begin{acknowledgements}
We thank Bob Fosbury, Reynier Peletier and Jose Acosta for
the very constructive comments which have helped improve this
paper. We are grateful to ST-ECF for generous allocations of computer
resources and to Richard Hook for its frequent assistance. MVM thanks also the
IAC
(Tenerife) for allocations of computer
resources. LB thanks the Observatoire de Lyon and STScI
for its hospitality, and aknowledges support from NASA
grants NAGW-3268 and GO-3724. MVM  aknowledges support from the Deustche
Forschungsgemeinschaft.
\end{acknowledgements}

\vspace{1.2cm}


\begin{appendix}
\section{Appendix}

By comparing the estimated photoionization rates from the excited
level 3d of Ca$^{+*}$, $\Pi_{3d}$, to that from the ground state 4s
of Ca$^+$, $\Pi_{4s}$, we show that photoionization of excited
Ca$^{+*}$ is a negligible process.

 	A fundamental parameter which determines the importance of
the ionization rate from the metastable level is its relative population
with respect the ground state: $\frac{n_{3d}}{n_{4s}}$, being
$n_{4s}$ the density of Ca$^+$ ions in the ground level and $n_{3d}$
the density of Ca$^+$ ions in the metastable level, 3d. To evaluate this
fraction,
we have solved analytically the statistical equilibrium equations.
To simplify the calculations  we have considered a three level atom, reducing
the
two 4p sublevels (see Fig.{}~1) to a single level and the same for the 3p
sublevels.
We have taken into account all the processes (collisional and radiative) which
can
populate or depopulate  each of the levels.

The resolution of the system of three equations gives us  the ratios
$\frac{n_L}{n_{Ca^+}}$, with L=3d,4s,4p, being $n_{Ca^+}$ the total
density of Ca$^+$ ions. The density and temperature we considered were
300$cm^{-3}$ and 10000K, respectively.  From this we deduced the
relative population with respect the ground state. The results turn
out to be:

$$\frac{n_{3d}}{n_{4s}} \sim 10^{-7}$$  and	$$\frac{n_{4p}}{n_{4s}} \sim 0$$

	The negligible population of the upper 4p level prevents any
contribution by cascade to the population of the 3d level, therefore
collisional excitation is the only important mechanism populating the
metastable 3d level. This is consistent with the fact that we do not
observe the triplet of Ca$^+$ (8498,8542,8662) (4p to 4s) in any EELR
although it is observed in the broad line region of AGN where
densities are higher by many order of magnitudes.

\vspace{0.2cm}

{\bf a) Ionization of Ca$^{+*}$ by soft continuum photons.}

\vspace{0.2cm}

The soft UV counterpart of the ionizing continuum provides a source of
ionization for both Ca$^{+}$ (IP: 11.9eV) and Ca$^{+*}$ (IP: 10.2eV). To
estimate the photoionization rates, we will make the following approximaptions:

\noindent 1) At the fairly large depth in the cloud where the specie
Ca$^+$ becomes abundant, we only need to consider  photons with
energies $<13.6eV$, the ionization potential (IP) of H$^0$, because photons
just above this energy have already been absorbed and also because of
the rapid decrease of the photoionization cross section with
increasing energies.

\noindent 2) The ionizing continuum which reaches the PIZ is considered to
be the soft UV counterpart of an ionizing PL of index $-1.4$ (but
unattenuated since the opacity due to dust or trace elements below
13.6eV is relatively small). The continuum impinging the cloud is
described by $F_\nu = F_s \nu^{-1.4}$. In number of photons this is
$F_{\nu}/h\nu = F_s/h \nu^{-2.4}$. The constants $F_s/h$ will cancel out
when taking the ratio $\Pi_{3d}/\Pi_{4s}$.

\vspace{0.2cm}

	The atomic data was taken from Osterbrock (1987) for H$^0$ and
from Shine \& Linsky (1975) for Ca$^+$ and Ca$^{+*}$. Tables 1, 2, 3
where we define $a'_{\nu}$ = $a_{\nu}*10^{18}$ and $\nu'=\nu/10^{16}$
show the repevant atomic data.  The threshold ionizing frequency of
H$^0$ is labelled $\nu_0$.

We  estimate the  quotient $\Pi_{3d}/\Pi_{4s}$ as follows:

$$\frac{\Pi_{3d}}{\Pi_{4s}}=
\frac{n_{3d}}{n_{4s}}
\frac{\int\limits_{\nu_{3d}}^{\nu_{0}}{\nu^{-2.4}a_{\nu}(Ca^{+*})
d\nu}}{\int\limits_{\nu_{4s}}^{\nu_0}{\nu^{-2.4}a_{\nu}(Ca^+) d\nu}}
{}~{}~(eq.1)$$
If we define $a'_{\nu}$ = $a_{\nu}*10^{18}$ and $\nu'=\nu/10^{16}$, we have

$$\frac{\Pi_{3d}}{\Pi_{4s}}=
\frac{n_{3d}}{n_{4s}}
\frac{\int\limits_{\nu'_{3d}}^{\nu'_{0}}{\nu'^{-2.4}a'_{\nu}(Ca^{+*})
d\nu'}}{\int\limits_{\nu'_{4s}}^{\nu'_0}{\nu'^{-2.4}a'_{\nu}(Ca^+) d\nu'}} $$

$$= \frac{n_{3d}}{n_{4s}} \frac{I_1} {I_2} {}~{}~(eq.2)$$
The values of $a'_{\nu}(Ca^{+*})$ at $\nu'_0, \nu'_{4s}$ and
$a'_{\nu}(Ca^{+})$ at $\nu'_0$ and $\nu'_{4s}$ have been obtained by
interpolation from (see Tables 2 and 3).
Using the rule of the rectangle to approximate a given integral:

{\small $$\int\limits_{x_0}^{x_1}f(x) dx \sim (x_1-x_0)f(x_0) +
(x_2-x_1)f(x_1) + ... $$
$$ + (x_n-x_{n-1}) f(x_{n-1})$$}

\begin{table}
\centering
\caption{$H^0$ photoionization cross section}
\begin{tabular}{lll} \hline
  $\nu'$ & $a'_{\nu}(H^0)$  & $\nu^{'-2.4}a'_{\nu}(H^0)$ \\ \hline
         0.3    &       6.4    &     115.1	 \\
         0.4    &       2.5     &    22.54	 \\
         0.6     &     1.15    &     3.919	 \\
         0.8     &      0.6    &     1.025	 \\
           1     &      0.3      &     0.3	 \\
         1.2     &      0.2    &    0.1291	 \\
         1.4     &     0.15   &    0.06689	 \\
         1.6    &       0.1  &     0.03237	 \\
         1.8    &      0.07  &     0.01708	 \\
           2     &     0.05   &   0.009473	 \\
         2.2    &      0.04  &    0.006029	 \\
         2.4    &      0.03   &     0.00367	 \\
         2.6    &      0.02   &    0.002019	 \\
         2.8    &     0.015  &    0.001267	 \\
           3    &     0.015   &   0.001074	 \\
         3.2    &      0.01  &   0.0006133	 \\
         3.4    &      0.01  &   0.0005302	 \\
         3.6    &      0.01  &   0.0004622	 \\

\end{tabular}
\end{table}

\begin{table}
\centering
\caption{Ca$^+$ 4s photoionization cross section}
\begin{tabular}{llll} \hline
$\lambda(\AA)$ &  $\nu'$ & $a'_{\nu}(Ca^{+})$  & $\nu^{'-2.4}a'_{\nu}(Ca^{+})$
\\ \hline
        1044 &     0.2873 ($\nu'_{4s}$) &     0.2036 &      4.063	\\
        1000 &        0.3 &     0.2097 &      3.772	\\
         950 &     0.3158 &     0.2145 &      3.412	\\
	  &    0.3288 ($\nu'_0$) &     0.2157 &      3.114     \\
         900 &     0.3333 &     0.2170 &      3.031	\\
         850 &     0.3529 &     0.2172 &      2.644	\\
         800 &      0.375 &     0.2149 &      2.262	\\
         750 &        0.4 &     0.2103 &      1.896	\\
         700 &     0.4286 &     0.2033 &      1.554	\\
         650 &     0.4615 &     0.1942 &      1.242	\\
         600 &        0.5 &     0.1830 &      0.966	\\
         550 &     0.5455 &     0.1700 &     0.7282	\\
         500 &        0.6 &     0.1554 &     0.5295	\\
         450 &     0.6667 &     0.1394 &      0.369	\\
         400 &       0.75 &     0.1225 &     0.2443	\\
         350 &     0.8571 &     0.1049 &     0.1518	\\
         300 &          1   &   0.0870  &   0.08697	\\
         250 &       1.2  &     0.0691   &   0.0440	\\
         200 &        1.5  &    0.0519 &    0.0199	\\
         150 &          2  &    0.0352 &    0.00668	\\
         100 &          3 &      0.0199 &   0.001417	\\
          50 &          6  &    0.0198  & 0.0002685	\\

\end{tabular}
\end{table}

\begin{table}
\centering
\caption{Ca$^{+*}$ 3d photoionization cross section}
\begin{tabular}{llll} \hline
$\lambda(\AA)$ & $\nu'$ & $a'_{\nu}(Ca^{+*})$  & $\nu^{'-2.4}a'_{\nu}(Ca^{+*})$
\\ \hline
       1218 &     0.2462 ($\nu'_{3d}$) &      6.148 &      177.70  \\
        1200 &       0.25 &      6.086 &      169.50  \\
        1150 &     0.2609 &      5.907 &      148.60	\\
        1100 &     0.2727 &      5.716 &      129.20	\\
        1050 &     0.2857  &      5.511 &      111.4	\\
 	1044 & 	    0.2873 ($\nu'_{4s}$)  &    5.403  &    107.80  \\
        1000 &        0.3 &      5.295 &      95.22	\\
         950 &     0.3158 &      5.066 &      80.56	\\
	    &     0.3288 ($\nu'_0$) & 4.9465 &  71.39  \\
         900 &     0.3333 &      4.827 &      67.41	\\
         850 &     0.3529 &      4.576 &      55.72	\\
         800 &      0.375 &      4.315 &      45.43	\\
         750 &        0.4 &      4.044 &      36.46	\\
         700 &     0.4286 &      3.762 &      28.74	\\
         650 &     0.4615 &       3.47 &      22.19	\\
         600 &        0.5 &      3.168 &      16.72	\\
         550 &     0.5455 &      2.856 &      12.23	\\
         500 &        0.6 &      2.534 &      8.634	\\
         450 &     0.6667 &      2.202 &      5.827	\\
         400 &       0.75 &      1.862 &      3.714	\\
         350 &     0.8571 &      1.517 &      2.196	\\
         300 &          1 &      1.173 &      1.173	\\
         250 &        1.2 &     0.8424 &     0.5439	\\
         200 &        1.5 &     0.5401 &     0.2041	\\
         150 &          2 &     0.2865   &    0.05428 	\\
     	 100 &  	3   &     0.1051   &   0.007524	\\
          50  &         6    &   0.01471  &   0.0001995 \\

\end{tabular}
\end{table}

(eq.2) reduces to

$$\frac{\Pi_{3d}}{\Pi_{4s}} \sim \frac{n_{3d}}{n_{4s}} \frac{ 10.06 }
{0.156}{}~{}~(eq.3)$$
If we take into account that $\frac{n_{3d}}{n_{4s}}\sim 10^{-7}$ we obtain

$$\frac{\Pi_{3d}}{\Pi_{4s}} \sim 6\times10^{-6}$$ which means that the
ionization of Ca$^+$ into Ca$^{++}$ originates overwelmingly from the
ground level 4s.

\vspace{0.2cm}

{\bf b) Ionization of Ca$^{+*}$ by Ly$\alpha$ photons.}

\vspace{0.2cm}

Ly$\alpha$ photons have energies slightly higher than the treshold
energy of level 3d of Ca$^{+*}$ and constitute undoudtedly  an important source
of
ionization for this level.  To compute the Ly$\alpha$ emissivity, we
assume the low density regime whereby $\simeq \frac{2}{3}$ of
recombinations of H$^+$ lead to the emission of a Ly$\alpha$ photon.
Since Ly$\alpha$ is a resonnant line of large line scattering opacity,
we consider that the density of Ly$\alpha$ photons within the Ca$^+$
region result from the production of Ly$\alpha$ either locally or from
{\it deeper} regions.  The justification for this is that resonnant line
photons generated from layers nearer the slab'surface would be
reflected outward as a result of the increasing fraction of H$^o$ with
depth (see Binette et{}~al 1993b).  Therefore the Ly$\alpha$ flux
potentially available to ionize level 3d is a small fraction $\eta$ of
the total Ly$\alpha$ flux emitted by the cloud. This fraction is of
order 0.2 corrresponding to the fraction of ionizing photons of H$^0$
{\it not yet} absorbed at the typical depth where the Ca$^+$ specie is
abundant.

	We first compute the ${\Pi^{Ly\alpha}_{3d}/\Pi_{4s}}$ ratio
taking into account that resonnant scattering will increase the
density of locally emitted line photons by a factor $\xi \sim 10^{7}$,
the mean number of scatterings before escape. Such a high number of
line scattering, however, characterizes only the locally produced
Ly$\alpha$ photons. This accumulation (or slowing down) effect which
we want to estimate is only effective within a zone of optical depth
of order a few. Let's take $\tau_{scat} \sim 10 = \sigma_{scat} n(H^0)
{}~\delta X$ where $\delta X$ is the geometrical depth and $n(H^0)$ the
local density of $H^{0}$.  Adopting $\sigma_{scat} \simeq 6 \times
10^{-14} cm^{2}$ (cf Appendix{}~B of Binette et{}~al 1993b), the column
density of recombining H$^+$ we should consider in generating
Ly$\alpha$ is within a thickness $N(H^+) \sim N(H^0)= n(H^0) {}~\delta X
= 10/6\times10^{-14} = 1.7\times 10^{14}cm^{-2}$.

	Taking these considerations into account, the problem reduces
to estimating the quotient

$$ \frac{\Pi^{Ly\alpha}_{3d}}{\Pi_{4s}}=\frac{n_{3d}}{n_{4s}}
\frac{\frac{2}{3} \eta \xi
a_{Ly\alpha}(Ca^{+*}) N(H^+)
\int\limits_{\nu'_0}^{\infty}{\nu'^{-2.4}a'_{\nu}(H^0)
d\nu'}}{\int\limits_{\nu'_{4s}}^{\nu'_0}{\nu'^{-2.4}a'_{\nu}(Ca^+) d\nu'}}$$
$$= \frac{n_{3d}}{n_{4s}} \frac{2}{3} \eta \xi a_{Ly\alpha}(Ca^{+*}) N(H^+)
\frac{I_3}{I_2}
{}~{}~ (eq.4)$$
where $a_{Ly\alpha}($Ca$^{+*})$ is the photoionizatiom cross
section of Ca$^{+*}$ at the Ly$\alpha$ energy ($\simeq 6\times10^{-18}
cm^{2}$).
Taking again ${n_{3d}}/{n_{4s}} \sim 10^{-7}$, we obtain

$$ \frac{\Pi^{Ly\alpha}_{3d}}{\Pi_{4s}}\sim 1.36 \times 10^{-4} \frac{I_3}{I_2}
{}~{}~ (eq.5)$$
Using again the rule of the rectangle we obtain

$$ \frac{\Pi^{Ly\alpha}_{3d}}{\Pi_{4s}}\sim 1.5\times10^{-2}.$$

	The effect of slowing down of resonnant line photons is
interesting but appears insufficient as it envolves too small a
fraction of the Ly$\alpha$ photons generated within the nebula. Let's
now estimate the importance of all the Ly$\alpha$ photons generated
within the deeper zones which, after having scattered far enough in
frequency to escape, must still cross the Ca$^+$ zone (without
appreciable scattering in that zone). The effect on the
photoionization of level 3d is given in this case by

$$ \frac{\Pi^{Ly\alpha}_{3d}}{\Pi_{4s}}=\frac{n_{3d}}{n_{4s}}
\frac{\frac{2}{3} \eta
a'_{Ly\alpha}(Ca^{+*}) \int\limits_{\nu'_0}^{\infty}{\nu'^{-2.4}
d\nu'}}{\int\limits_{\nu'_{4s}}^{\nu'_0}{\nu'^{-2.4}a'_{\nu}(Ca^+)
d\nu'}} {}~{}~ (eq.6)$$
$$= \frac{n_{3d}}{n_{4s}} \frac{2}{3} \eta  a'_{Ly\alpha}(Ca^{+*})
\frac{I_4}{I_2}
{}~{}~(eq.7)$$
(Note that $a'$ is used here for Ca$^{+*}$). Using the rule of the rectangle
we obtain

$$ \frac{\Pi^{Ly\alpha}_{3d}}{\Pi_{4s}}\sim 2.43\times10^{-6}$$

In summary, Ly$\alpha$ emission and/or trapping are insufficient to
ionize Ca$^+$. As in the case of soft continuum photons, this is
basically the result of the extremely small population characterizing
the excited level 3d.

\end{appendix}

\end{document}